\documentclass{article}
\usepackage{apacite}
\usepackage{setspace}
\usepackage{xcolor}
\usepackage{siunitx}
\usepackage{bigints}
\usepackage{changepage}
\usepackage{graphicx}
\graphicspath{ {Images} }
\DeclareUnicodeCharacter{2009}{\,}

\usepackage[legalpaper, portrait, margin=1in]{geometry}
\title{\textbf{Multi-Messenger Observability of Neutron Star Binary Systems}}
\author{Jeshwanth Mohan\\}
\date{}

\begin{document}
\UseRawInputEncoding
\maketitle

\begin{center}
    \section*{Abstract}
\end{center}

    \begin{adjustwidth}{}{}
       \textrm{As technology has improved, binary neutron star systems have been observed more frequently, in fact, the first gravitational wave to have an electromagnetic counterpart  originated from the merger of two neutron stars (GW170817). Detecting these systems prior to merger may help recover essential data for developing an Equation of State for neutron stars. This paper examines the observability of detached eclipsing binary neutron stars prior to merger by simulating the potential observability of neutron star systems in the optical. It is found that it is not likely considering current instruments due to low visibility and inadequate time resolution, however, improvements in the future or a wide field X-ray instrument may offer a viable option for detecting these systems.}
    \end{adjustwidth}

\section{Introduction}
    \textrm 
    {The binary star merger preceding gravitational waves (GW) along with electromagnetic signals across the spectrum \textcolor{blue}{\cite{Gamma}}, has led to a leap toward understanding the nature of gravitational waves. This multi-messenger event in 2017 included the emission of gravitational waves (GW170817) \textcolor{blue}{\cite{GW}}, gamma ray bursts (GRB170817A), and a relativistic jet \textcolor{blue}{\cite{Gamma}}. This event was the first direct detection of gravitational waves, allowing scientists to observe the event through both electromagnetic and gravitational methods. The observation of such an event may allow scientists to further explore and concretely define an Equation of State (EOS) for neutron stars. 
    \\ \indent{} This event and all similar events can be split into three different stages: pre-merger, during the merger, and post-merger. Before the merger, the pre-merger stage, the two neutron stars are in an eclipsing and detached system. An eclipsing system indicates two stars orbiting each other in a way that they overlap each other in the plane of view, decreasing the total light observed from the system. A detached system indicates that they are within their respective Roche lobes \textcolor{blue}{\cite{Roche}}, the area around a star where matter is gravitationally bound. This prevents mass transfer from either star. As the two stars near each other in their collapsing orbit, they begin to create ripples in the space-time curvature. These ripples are the gravitational waves that many instruments detected in 2017. During the merger, the neutron stars collide and release enormous amounts of energy and reaching bolometric absolute magnitudes of -21 or \num{5e26} times brighter than the sun \textcolor{blue}{\cite{1998}}. These events are called kilonovae, which are one of the sources for many of the heavier elements such as gold \textcolor{blue}{\cite{2013}}. Despite being very bright these events are short-lived making their detection harder. The energy released ranges from radio to gamma on the electromagnetic spectrum, however, with the majority of the energy being emitted through the form of X-ray and short gamma ray bursts (SGRBs) \textcolor{blue}{\cite{Gamma}}. The electromagnetic signals are accompanied by gravitational waves. Lastly, the post-merger stage is the remnant of the kilonova, either a low-mass black hole (BH) or a neutron star (NS). These remnants may not hold as much information as their progenitors, nevertheless, they hold vital information.
    \\ \indent{} Due to the lack of prior detection, the progenitors of this multi-messenger event are relatively unknown. However, estimates for their masses have been estimated by the LIGO Scientific Collaboration and Virgo Collaboration. The masses were estimated using a GW Parameter Estimation \textcolor{blue}{\cite{Proginetor}}. To further our understanding of such events, efforts have been taken to model these systems and understand the propagation of such electromagnetic and gravitational waves. Detection prior to the collision may allow for key data to understand the radial velocity and mass relationship of neutron stars.
    \\ \indent{} In order to improve the understanding of the observability of binary neutron star (BNS) systems, this study models and simulates the progenitors of GW170817 to characterize potential BNS systems. 
    \\ \indent{} This paper is organized with Section \textcolor{blue}{2} describing the methods, Section \textcolor{blue}{3} disclosing and analyzing the results, and Section \textcolor{blue}{4} discussing the results with relation to the current state of understanding.}

\hspace{-0.75in} \includegraphics[width = 7.5in]{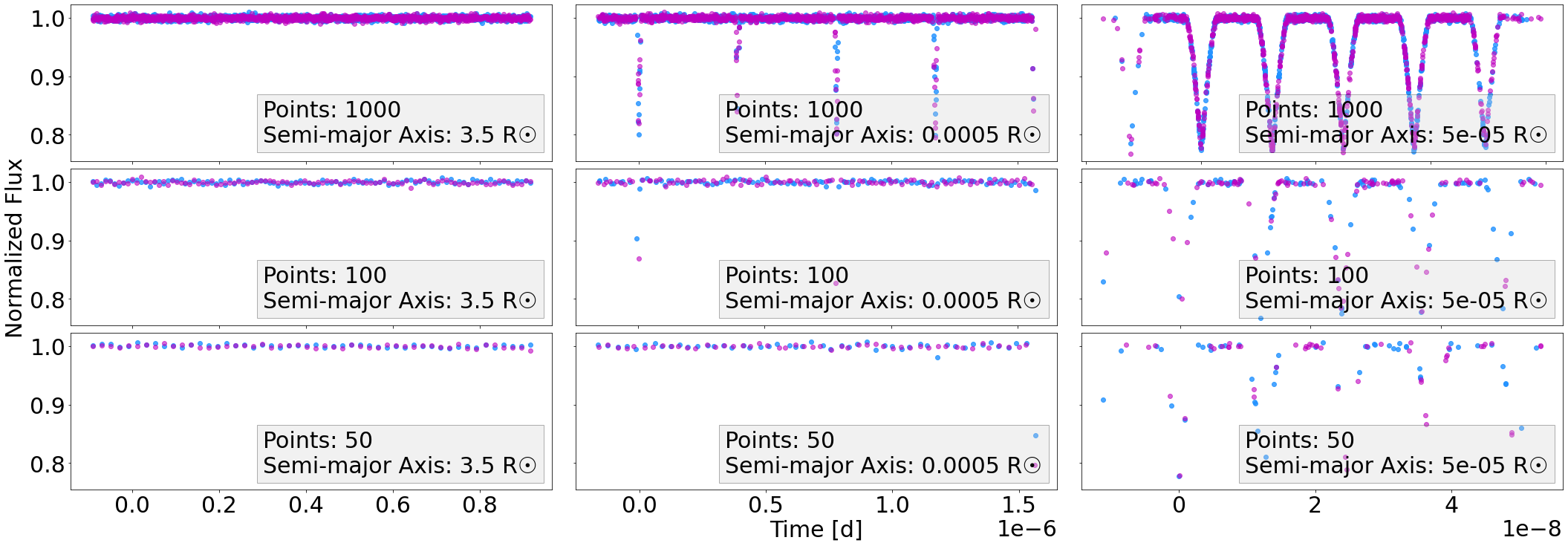} \textrm{\textbf{Figure 1.} Grid view of 9 different simulated light curves of a binary neutron star system tested at different semi-major axis (3.5$R_\odot$, \num{5e-4}$R_\odot$, and \num{5e-5}$R_\odot$) and  different time resolutions (1000 points, 100 points, and 50 points).}

\section{Methods}
    \textrm
    {To model the BNS system, the radii and masses of the stars are necessary. This study utilizes the predicted masses of the progenitors of GW170817 to complete all necessary simulations. Data from this event is used as it is the most widely available and accurate information to use for simulation. A paper on the progenitors of this event by the LIGO Scientific Collaboration and Virgo Collaboration presents the range of the masses to be 1.36-1.60$M_\odot$ and 1.17-1.36$M_\odot$ \textcolor{blue}{\cite{Proginetor}}. For the purposes of this paper, the masses used were averages of the ranges given, 1.48$M_\odot$ and 1.26$M_\odot$ respectively \textcolor{blue}{\cite{Proginetor}}. Estimations of the radii are 10.8 km for the more massive star and 10.7 km for the less massive star \textcolor{blue}{\cite{EOS}}. This data will be used to generate simulated light curves and to construct a possible orbit. There are three main aspects to this study, the generation of light curves, the recovery of period, and integration of the orbit, all these will be discussed in detail in the following subsections.}

    \subsection{Light Curves}
        \textrm
        {To generate light curves, the publicly available code for ELLC, version 1.8.6, was used \textcolor{blue}{\cite{ELLC}}. This code offers a method of graphing the flux or luminosity of a binary star system using the radii, semi-major axis, surface brightness ratio, mass ratio, and inclination. A constant surface brightness ratio and mass ratio were used as these were intrinsic properties of the progenitors. The inclination was tested  near 90$^{\circ}$ to allow for maximum observability. To simulate the collapsing orbit, the semi-major axis was varied from the lower limit, the Roche limit, and the upper bound the predicted original axis, which is estimated to around 3.5$R_\odot$ \textcolor{blue}{\cite{Proginetor}}.}
        \\
        
    \begin{center}
        \includegraphics[width = 0.49\linewidth]{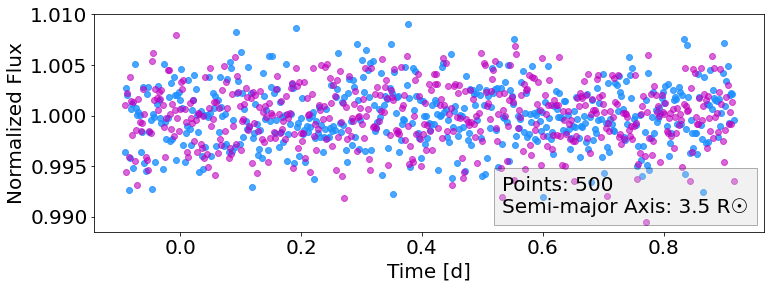}
        \includegraphics[width = 0.49\linewidth]{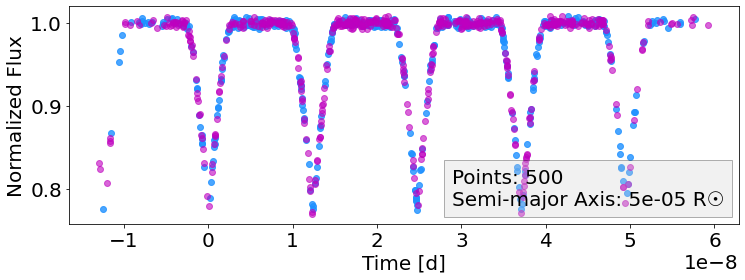}
    \end{center}
    \textrm{\textbf{Figure 2.} Two light curves simulating the progenitors of GW170817. The left panel represents the system prior to in-fall. The right panel depicts the system right before the merger.}
    
    \subsection{Period Recovery}
        \textrm{Once the light curves generated they were analyzed by the GATSPY software which returned a period to best fit the curve \textcolor{blue}{\cite{2015}}. To be considered an accurate period, the returned value should either be two times, one-half, or almost exactly the actual period which is found using Kepler's Laws. This is done because while analyzing the data, the software may fold the data causing it to return half or 2 times the period. The software folds the period in order to find the repeated cycles, however, when the dips are extremely similar, as in this situation, the software cannot discern between the different dips and returns half of the period. In this study, the neutron stars are extremely similar in terms of luminosity and size, thus, they cause similar dips causing the software to always return half of the period (see Figure \textcolor{blue}{3}). In Figure \textcolor{blue}{3}, the recoverable period falls in the half period range. Another peak at one-fourth of the actual period is also observed, which indicates the similarity of the dips in flux causing the software to fold two times more than necessary. Simulations that yielded periods that follow the description above can be classified as recoverable. }
        
    \begin{center}  
        \includegraphics[width = 0.49\linewidth]{/500_5e-5.png}
        \includegraphics[width = 0.49\linewidth]{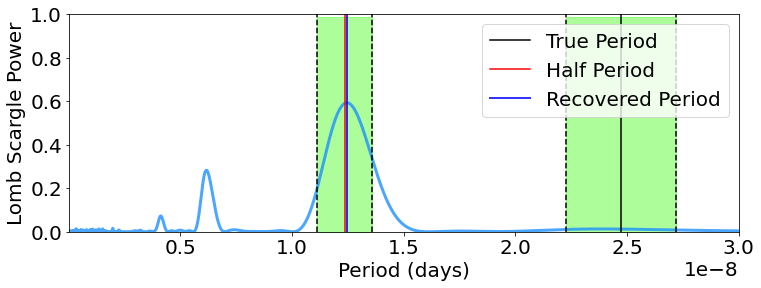}
    \end{center}
    \textrm{\textbf{Figure 3.} The right panel is a periodogram of the light curve from the left panel. The blue curve indicates the probability of the likelihood of each period. The black and red line on the right panel depict the true and half period. The shaded green areas show acceptable ranges for the recovered period.}

    \subsection{Orbit Integrator}
        \textrm
        {To further understand the situation leading up to the event, this study includes an orbit integrator which maps out a possible path taken by the two eclipsing stars. A Runge-Kutta 4$^{th}$ order method to solve the differential equations necessary to generate the positional and velocity vectors of the two bodies in the simulation.  
        }
        
\begin{center}
    \includegraphics[width=0.99\textwidth]{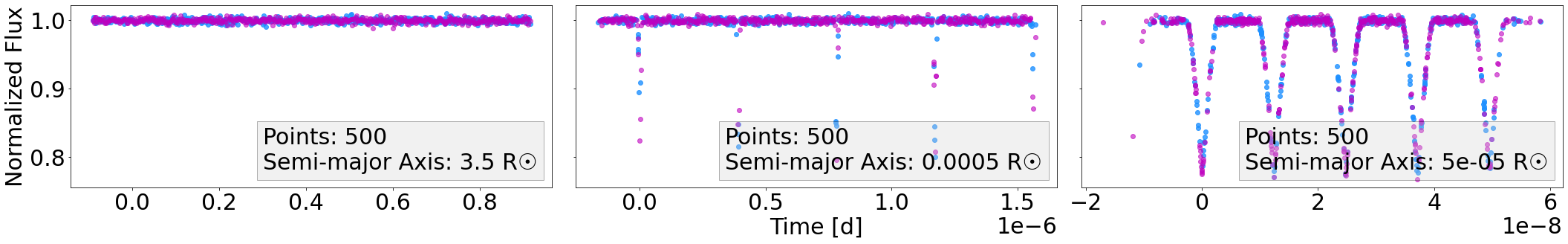}
    \includegraphics[width=0.329\textwidth]{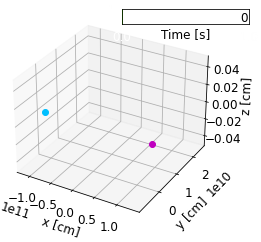}
    \includegraphics[width=0.329\textwidth]{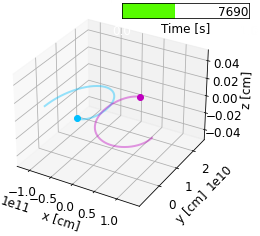}
    \includegraphics[width=0.329\textwidth]{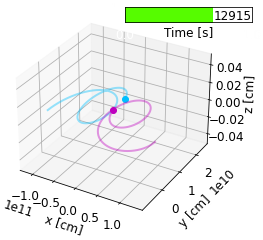}
\end{center}

\begin{adjustwidth}{}{}
\textrm{\textbf{Figure 4.} \textrm{Above are snapshots of the orbit integrator and the corresponding light curves at the distances of 3.5 AU, \num{5e-4} AU, and \num{5e-5} AU.The animation of the complete orbit can be found here: \url{https://drive.google.com/file/d/1CNV6KFfawy3PGEMM3dIPPNDjtcjv-M3s/view?usp=sharing}}}
\end{adjustwidth}

\section{Observability}
    \textrm
    {Studies on neutron stars are limited and have not completely understood their mechanisms making them very mysterious objects. This is partially due to their low visibility to current instruments.}
    
    \subsection{Preliminaries}
    
    \textrm{The luminosity of neutron stars can be described by the Stefan-Boltzmann law, which relates the luminosity to the temperature and the radius, which is given as $L = 4 \pi R^2\sigma T^4$. Using the average radius and surface temperature of a neutron star, this equation leads to a luminosity of about 0.2$L_\odot$ or an absolute magnitude of 6.58. }
    \\ \indent{} 
    \textrm {This bolometric luminosity encompasses all wavelengths of electromagnetic radiation and thus is much brighter than what could be detected. To find what is actually observable in a specific wavelength or frequency it is necessary to perform a K-correction. Which is given as }
    
    \[ m_r = M_Q + DM + K_{QR} \]
  
    \textrm{To find the actual K-correction it is necessary to integrate the spectral flux density from a specific range. The full equation \textcolor{blue}{\cite{K-correction}} is defined as}
    
    \[ K_{QR} = -2.5\log_{10} \left[ \left[ 1 + z \right] \frac{\bigintsss \frac{d\nu_o }{\nu_o} L_{\nu}(\left[ 1 + z \right]\nu_o)R(\nu_o) \bigintsss \frac{d\nu_e}{\nu_e} g_v^Q(\nu_e) Q(\nu_e)}  {\bigintsss \frac{d\nu_o}{\nu_o} g_v^R(\nu_e) R(\nu_o) \bigintsss \frac{d\nu_e}{\nu_e} L_v(\nu_e) Q(\nu_e)} \right] \]
    
    \textrm {This equation accounts for redshift and differences between band passes, however, all of this has no effect in this case since the possible distances observed by current technology are extremely local, thus simplifying into an integral bounded by a specific range of frequency of electromagnetic waves over the total flux. The flux density for a black body is defined by Planck's Law \textcolor{blue}{\cite{Planck}}:}

    \[  B(\nu,T) = \frac{2h\nu^3}{c^2}\frac{1}{e^{\frac{h\nu}{k_BT}} - 1} \]
    
    \textrm{Where $h$ is Planck's constant, $c$ is the speed of light, $k_B$ is the Boltzmann constant, $T$ is the temperature, and $\nu$ is the frequency. Utilizing this, $K_{QR}$ can be calculated with the following expression}
    
    \[  -2.5\log_{10} \left[ \frac{\bigints_{\nu_1}^{\nu_2} \frac{2h\nu^3}{c^2}\frac{1}{e^{\frac{h\nu}{k_BT}} - 1}}{\sigma T^4} \right]  \]
    
    \textrm{For this study two bands will be used: visual and X-ray. The frequency for the visual band ranges from \num{4e14} Hz to \num{8e14} Hz. X-ray ranges from 0.1 keV to 10 keV or \num{2.4e16} Hz to \num{2.4e18} Hz. Inputting these bounds into the formula above returns a $K_{QR}$ of 15.25 for the visual band and 1.30 for the X-ray band. This is accurate because the majority of the energy released from a neutron star would be in the X-ray region as described by Wien's Displacement law. Finally, these K-corrections found can be used to manipulate the K-corrected distance modulus formula to find the farthest distance a telescope can detect such objects.}
    
    \[ D = 10^{\frac{m_r - M_Q - K_{QR}}{5} + 1} \]
    
    \textrm{Beyond that, to observe these eclipsing binaries it would be necessary to detect the slight dip in the total flux, which in this case was around 25\%. This indicates a dip of 0.05$L_\odot$ which would indicate a change of 0.31 in the apparent magnitude which would be added to the $K_{QR}$ making it 15.56 and 1.61 for the visual and X-ray bands respectively.}

    \subsection{Limitations}
    \textrm{Current technology is limited by its lower and upper limits on observable magnitude. For example, the ASAS-SN sky patrol mainly detects between the apparent magnitude of 9 to 18 in the visual band \textcolor{blue}{\cite{ASASSN}}. The lower limit can be inputted into the modified distance modulus formula to find that the farthest distance such eclipsing binaries could be detected is 1.49 parsecs. Another example is the Zwicky Transient Facility \textcolor{blue}{\cite{ZTF}} which has a 5$\sigma$ limiting magnitude of 20.8 in the visual band resulting in a detection distance of 5.40 parsecs. Unlike the last two observatories, the Vera C. Rubin Observatory \textcolor{blue}{\cite{Rubin}} has a 5$\sigma$ limiting magnitude of 24.82 in the visual band which provides a much farther range of detection of 34.36 parsecs.  }
    
    \textrm{Utilizing these distances, an observable volume can be determined in order to predict how many neutron star mergers may be detected. For GW170817A type mergers the density of these mergers is around 300 Gpc$^{-3}$ yr$^{-1}$ \textcolor{blue}{\cite{Rates}}. The estimated merger frequencies achieved by using the distances are \num{4.15e-24}, \num{1.98e-22}, and \num{5.10e-20} mergers per year for each of the observatories described in that order. }

    \textrm {Furthermore, this study utilizes inclinations near 90$^\circ$ which is unlikely considering the vast number of orientations of a BNS system.}
    
    \textrm{However, it may be easier to detect these objects using X-ray. For example, the Chandra X-ray Observatory \textcolor{blue}{\cite{CHANDRA}} has a limiting flux of \num{3.33e-14} ergs s$^{-1}$ cm$^{-2}$ which translates to a limiting apparent magnitude of 22.34. This leads to a maximum distance of detection of 6754.6 parsecs. With a much larger volume, the observatory can detect \num{9.24e-5} mergers per year. Nonetheless, this is only applicable if a wide-field X-ray observatory is established, and even then the rate of observation would be extremely low.}

    \textrm{Thus, detecting these systems is currently almost impossible because our current technology struggles to detect the neutron stars themselves, plus, there are only fractions of seconds to capture the necessary data in order to recover the period. All these limitations make it difficult to expand current knowledge of these objects through direct observations, however, post merger data may still be used to study neutron stars.}
    
    \textrm{}
    
    \subsection{Post-Merger Detection}
    
    \textrm{Post-merger data can reveal important properties of the progenitors, but obviously direct observations hold much more accurate and concrete information. Nonetheless, post-merger data can be used for example the masses of the two neutron stars were derived from the kilonova light curves, while the gravitational waves offered a completely different perspective of viewing the universe. The left panel in Figure \textcolor{blue}{5} depicts the gravitational wave data from several different observatories. This "chirp" was accompanied by the kilonova electromagnetic emission (Right Panel Figure \textcolor{blue}{5}). }
    
    \begin{center}  
        \includegraphics[width = 0.40\linewidth]{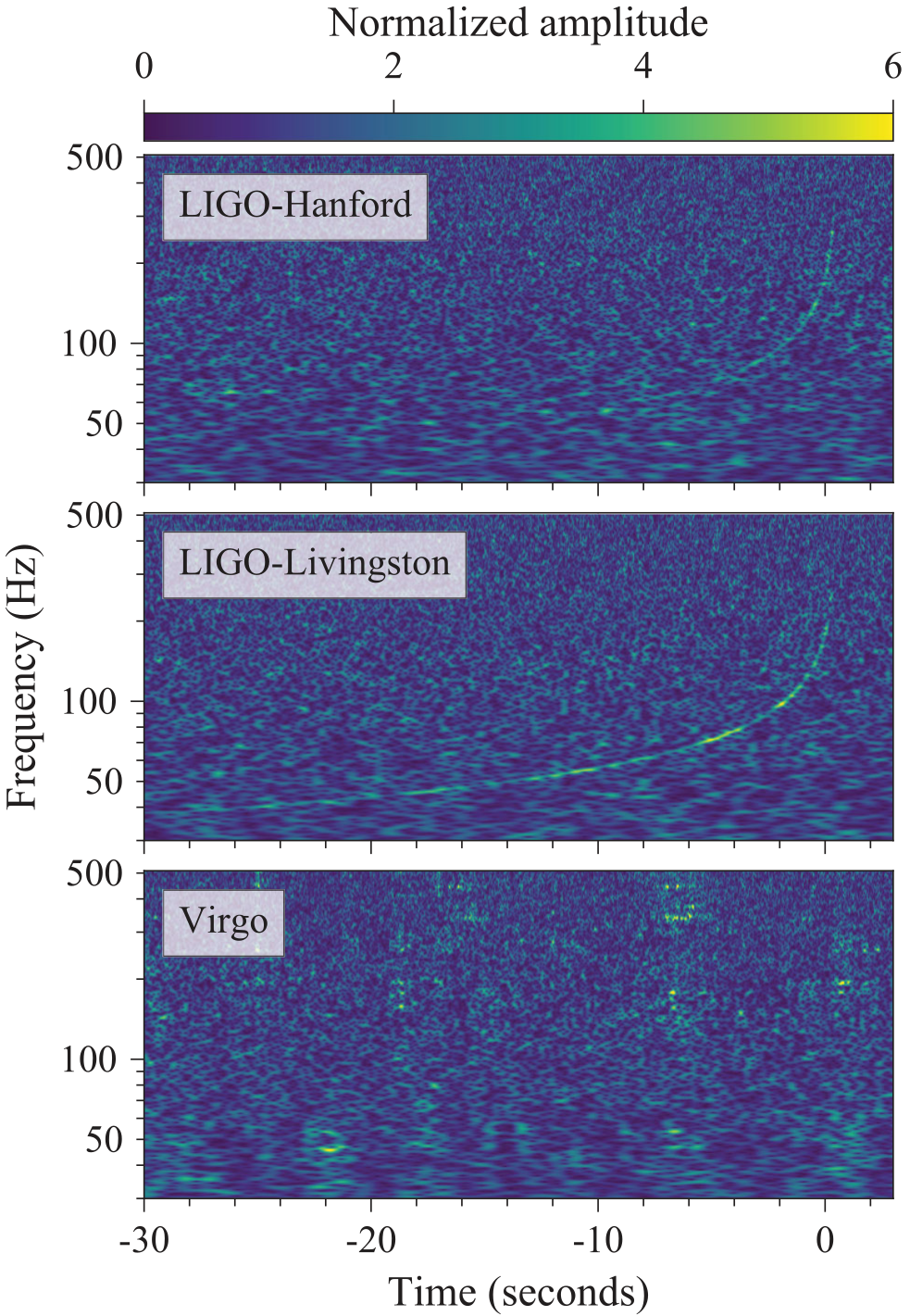}
        \includegraphics[width = 0.58\linewidth]{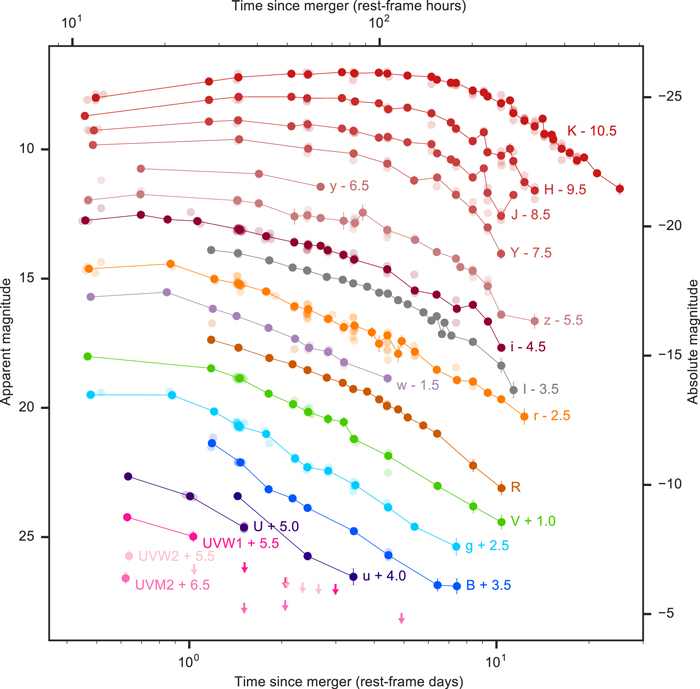}
    \end{center}
    
    \begin{adjustwidth}{}{}
     \textrm{\textbf{Figure 5.} The left panel is the data from gravitational event GW170817 \textcolor{blue}{\cite{Chirp}}. The right panel depicts the light curve of the kilonova resulting from the merger \textcolor{blue}{\cite{Kilonova}}}
    \end{adjustwidth}

\section{Conclusions}
    \textrm
    {This study simulates a neutron binary system based on the progenitors of GW170817 to understand detectability of such eclipsing binary stars. This study tests the observability of neutron stars at different points of the orbit and at different cadences. It is found that at the pre-merger orbit, the period is not recoverable because the neutron stars are extremely small and thus do not overlap each other enough to create an observable dip. However, as the in-falling system gets closer to collision the dips in the light curves become more prominent and thus the GATSPY software is capable of recovering the period. Furthermore, it is found that the period is recoverable at cadences of above 50 points over the course of several milliseconds indicating that an observatory must record data every 0.08 milliseconds while observing these systems in order to acquire enough data points to recover the period. This time resolution is currently very difficult to achieve in wide-field observatories. Detection of BNS systems becomes more difficult when sensitivity is considered; the majority of optical observatories cannot detect neutron stars beyond 50 parsecs, while the closest observed neutron star, RX J1856, is over 100 parsecs away. However, neutron stars emit more prominently in the X-ray band making them far more visible allowing for a theoretical wide-field X-ray observatory to detect neutron stars and BNS systems up to around 6750 parsecs away. Nonetheless, this is still not adequate for feasibly detecting BNS systems due to their scarcity in the universe. Thus, establishment of a wide-field X-ray observatory while simultaneously improving sensitivity may lead to detection of these systems in the future.}

\bibliographystyle{apacite}
\bibliography{references}

\end{document}